\documentclass[a4paper,11pt]{article}

\usepackage{eucal}
\usepackage{amsfonts,amsmath}
\usepackage{graphics,epsfig}
\usepackage{mathrsfs,latexsym}
\usepackage{fancyhdr}
\usepackage{cite}

\def\bbc{{\Bbb C}}
\def\bbr{{\Bbb R}}

\def\bbz{{\Bbb Z}}

\def\asl{\mbox{\rm sl\,}}

\def\re{\mathrm{Re\,}}
\def\im{\mathrm{Im\,}}

\def\Res{\mathop{\mbox{Res}\,}\limits}

\def\rmd{\mathrm{d\,}}
\def\rmi{\mathrm{i}}
\def\rme{\mathrm{e}}

\def\openone{\leavevmode\hbox{\small1\kern-3.3pt\normalsize1}}
\def\diag{\mbox{diag\,}}

\begin{document}
\title{Pseudo-Hermitian Reduction of a Generalized Heisenberg Ferromagnet Equation.
II. Special Solutions}
\author{T. I. Valchev$^1$ and A. B. Yanovski$^2$\\
\small $^1$ Institute of Mathematics and Informatics,\\
\small Bulgarian Academy of Sciences, Acad. G. Bonchev Str., 1113 Sofia, Bulgaria\\
\small $^2$ Department of Mathematics \& Applied Mathematics,\\
\small University of Cape Town, Rondebosch 7700,
Cape Town, South Africa\\
\small E-mails: tiv@math.bas.bg, Alexandar.Ianovsky@uct.ac.za}
\date{}
\maketitle

\begin{abstract}
This paper is a continuation of our previous work in which we studied a $\asl(3)$ Zakharov-Shabat type
auxiliary linear problem with reductions of Mikhailov type and the integrable hierarchy of nonlinear
evolution equations associated with it. Now, we shall demonstrate how one can construct special solutions
over constant background through Zakharov-Shabat's dressing technique. That approach will be illustrated
on the example of a generalized Heisenberg ferromagnet equation related to the linear problem for $\asl(3)$.
In doing this, we shall discuss the difference between the Hermitian and pseudo-Hermitian cases.
\end{abstract}

\section{Introduction}

In \cite{gmv}, the system of completely integrable equations
\begin{equation}\label{ghf_0}
\begin{split}
&\rmi u_t + u_{xx}+( uu^*_x +  vv^*_x)u_x
+ ( uu^*_x + vv^*_x)_x u = 0 \, ,\qquad \rmi=\sqrt{-1}\, ,\\
&\rmi v_t + v_{xx}+(uu^*_x + vv^*_x)v_x
+ ( uu^*_x + vv^*_x)_x v  = 0
\end{split}\end{equation}
and the corresponding auxiliary spectral problem were introduced and studied. Above, subscripts
mean partial differentiation, $*$ denotes complex conjugation and smooth functions $u,v: \bbr^2\to\bbc$
are subject to the condition $|u|^2+|v|^2 = 1$ ($|z| = \sqrt{zz^*}$, $z\in\bbc$). System (\ref{ghf_0})
is of particular interest since it is an integrable generalization of the classical integrable
Heisenberg ferromagnet equation
\[\mathbf{S}_t = \mathbf{S}\times \mathbf{S}_{xx}\]
for the unit spin vector $\mathbf{S} = (S^{\,1},S^{\,2},S^{\,3})$, see \cite{BoPo90, book,blue-bible} for more details.

Further detailed analysis of (\ref{ghf_0}) and its spectral problem was carried out in \cite{jgsp,side9,yan, yanvil}.
In \cite{jgsp,side9}, the authors described the hierarchy of nonlinear evolution equations (NLEEs) associated with
(\ref{ghf_0}), the hierarchies of conservation laws as well as the hierarchies of Hamiltonian structures in the case
when $u(x)\to \rme^{\rmi\delta_{\pm}}$ and $v(x)\to 0$ sufficiently fast as $x\to\pm\infty$. Moreover, the generalized
Fourier transform interpretation of the inverse scattering transform for (\ref{ghf_0}) was established and special
soliton-like solutions to (\ref{ghf_0}) were constructed via dressing method. A deep study of the properties of the
recursion operators for the system (\ref{ghf_0}) and the geometry linked to those was carried out in \cite{yan, yanvil}. 

In \cite{yantih}, the authors of the current paper considered a bit more general system of NLEEs having the form:
\begin{equation}
\begin{split}
&\rmi u_t + u_{xx}+(\epsilon uu^*_x +  vv^*_x)u_x
+ (\epsilon uu^*_x + vv^*_x)_x u  =  0\, , \qquad \epsilon^2 = 1 \, ,\\
&\rmi v_t + v_{xx}+(\epsilon uu^*_x + vv^*_x)v_x
+ (\epsilon uu^*_x + vv^*_x)_x v  = 0
\end{split} \label{ghf}
\end{equation}
where $u$ and $v$ satisfy the constraint:
\begin{equation}
\epsilon |u|^2 + |v|^2 = 1 \, .
\label{constr}\end{equation}
Obviously, system (\ref{ghf}) coincides with (\ref{ghf_0}) when $\epsilon = 1$ and much like it (\ref{ghf})
is completely integrable with a Lax pair given by:
\begin{eqnarray}
L(\lambda) & = & \rmi\partial_x - \lambda S, \qquad \lambda\in\bbc,\qquad
S = \left(\begin{array}{ccc}
0 & u & v\\ 
\epsilon u^* & 0 & 0 \\
v^* & 0 & 0\end{array}\right), \label{ghf_lax1}\\
A(\lambda) &=& \rmi\partial_t + \lambda A_1 + \lambda^2 A_2,
\qquad A_2  = \left(\begin{array}{ccc}
- 1/3 & 0 & 0 \\ 0 & 2/3 - \epsilon|u|^2 & -\epsilon u^*v \\
0 & - v^*u & 2/3 -|v|^2
\end{array}\right),\label{ghf_lax2}\\
A_1 & = & \left(\begin{array}{ccc}
0 & a & b\\ 
\epsilon a^* & 0 & 0 \\
b^* & 0 & 0\end{array}\right),\quad
\begin{array}{ccc} 
a & = & - \rmi u_x - \rmi\left(\epsilon uu^*_x + vv^*_x\right)u\\
b & = & - \rmi v_x - \rmi\left(\epsilon uu^*_x + vv^*_x\right)v
\end{array}\label{ghf_lax3}.
\end{eqnarray}
Following the convention in \cite{yantih}, the case when $\epsilon = 1$ will
be referred to as Hermitian while the case when $\epsilon = -1$ --- pseudo-Hermitian.
 
In \cite{yantih}, we described the integrable hierarchy associated with (\ref{ghf}) in terms of recursion
operators and derived completeness relations of their eigenfunctions. Instead of building the theory from
the "scratch", our analysis was based on the gauge equivalence between the auxiliary spectral problem with
$L$ in the form (\ref{ghf_lax1}) and the one for a nonlinear Schr\"odinger equation, i. e. a generalized
Zakharov-Shabat system. This permitted us to obtain all our results for arbitrary constant asymptotic values
of the potential functions appearing in the auxiliary linear problems. 

Another issue of fundamental importance concerns the solutions of (\ref{ghf}). We intend to show in the
present paper how one can derive particular solutions to (\ref{ghf}) in the simplest case of trivial
(constant) background. In doing this, we shall not make use of the gauge equivalence between
(\ref{ghf_lax1})--(\ref{ghf_lax3}) and a generalized Zakharov-Shabat's system in canonical gauge. Our
approach will be based on Zakharov-Shabat's dressing method that seems to be suitable for the system of
equations we are interested in. 

The paper itself is organized as follows. Next section contains our main results and it is divided into
four subsections. The first subsection is preliminary --- its purpose is to give the reader some general
idea of Zakharov-Shabat's dressing method and how it could be applied to (\ref{ghf}). After deriving all
general formulas, we are going to construct special types of solutions in the subsections to follow. It
turns out there exist three different cases:
\begin{itemize}
\item  Generic case, when the poles of dressing factor are complex numbers in general position, i.e.
those are not real or imaginary numbers;
\item The case, when the poles of dressing factor are imaginary;
\item Degenerate case, when the poles of dressing factor are real.
\end{itemize}
Each case is considered in a separate subsection. The first two cases lead to soliton type solutions while the degenerate
one leads to quasi-rational ones. As we shall see, there are essential differences between the properties of the solutions
obtained in the Hermitian and pseudo-Hermitian cases. For example, some soliton type solutions develop singularities
in the pseudo-Hermitian case while their counterparts are non-singular in the Hermitian one. Moreover, it turns out that
quasi-rational solutions are possible in the pseudo-Hermitian case only.

Last section contains some concluding remarks.

\section{Special Solutions}\label{solutions}

In this section, we shall present our main results: construction of special solutions to the system
(\ref{ghf}) over constant background. More specifically, we shall assume its solutions obey the following
boundary condition:
\begin{equation}
\lim_{x\to\pm\infty}u(x,t) = 0 \, ,\qquad \lim_{x\to\pm\infty}v(x,t) = 1
\label{triv_back}\end{equation}
that is easily seen to be compatible with (\ref{constr}). Our approach to obtain particular solutions
will be based on Zakharov-Shabat's dressing method \cite{ZS,zakh-mikh}. Since we aim at providing a
self-contained exposition, we shall remind the reader its basics as applied to (\ref{ghf}) and the Lax
pair associated with it.

\subsection{Dressing Method and Linear Bundles in Pole Gauge}\label{dressing}

Dressing procedure represents an indirect method to solve completely integrable equations,
i.e. one  constructs new solutions to a given NLEE from a known (seed) solution. In doing
this, one essentially uses the existence and the form of Lax representation.

To see how dressing method works, consider the Lax pair
\begin{eqnarray}
L_0(\lambda) & = & \rmi\partial_x - \lambda S^{(0)}, \qquad 
S^{(0)} = \left(\begin{array}{ccc}
0 & u_0 & v_0\\ 
\epsilon u^*_0 & 0 & 0 \\
v^*_0 & 0 & 0\end{array}\right), \label{lax1_bare}\\
A_0(\lambda) & = & \rmi\partial_t + \sum_{k=1}^{N}\lambda^k A^{(0)}_k ,
\qquad N\geq 2,\qquad \lambda\in\bbc\label{lax2_bare}
\end{eqnarray}
where $u_0$ and $v_0$ are two known functions subject to conditions (\ref{constr}) and
(\ref{triv_back}). For any fixed $N$ the zero curvature condition 
\begin{equation}
[L_0(\lambda), A_0(\lambda)] = 0
\label{comp}\end{equation}
leads to some NLEE belonging to the integrable hierarchy of (\ref{ghf}), see \cite{yantih}. Thus,
$u_0$ and $v_0$ are assumed to be known solutions to that NLEE which obey (\ref{constr}) and
(\ref{triv_back}). Similarly to (\ref{ghf_lax1})--(\ref{ghf_lax3}), the above Lax operators fulfill
the following symmetry conditions:
\begin{eqnarray}\label{red10}
HL_0(-\lambda)H & = & L_0(\lambda) \, ,\quad HA_0(-\lambda)H = A_0(\lambda) \, ,\quad
H=\diag(-1,1,1) \, ,\\ \label{red20}
Q_{\epsilon}\left(S^{(0)}\right)^{\dag}Q_{\epsilon} & = & S^{(0)},\qquad
Q_{\epsilon}\left(A^{(0)}_k\right)^{\dag}Q_{\epsilon} = A^{(0)}_{k},\qquad
Q_{\epsilon} = \diag(1,\epsilon,1)
\end{eqnarray}
where $\dag$ stands for Hermitian conjugation.

Let $\psi_0$ be an arbitrary fundamental solution to the auxiliary linear problem:
\begin{equation}
L_0(\lambda) \psi_0(x,t,\lambda) =  0 \, .
\label{bare1}
\end{equation}
Due to (\ref{comp}), $\psi_0$ also fulfills the linear problem:
\begin{equation}
A_0(\lambda)\psi_0(x,t,\lambda) = \psi_0(x,t,\lambda) f(\lambda)
\label{bare2}
\end{equation}
where
\begin{equation}
f(\lambda) := \lim_{x\to\pm\infty}\sum_{k=1}^{N}\lambda^k\hat{g}_0(x,t)A^{(0)}_k(x,t)g_0(x,t)
\label{disp_law}\end{equation}
is the dispersion law of NLEE. Above, 
\[g_0 := \frac{1}{\sqrt{2}}\left(\begin{array}{ccc} 1&0&-1\\ \epsilon u^*_0&\sqrt{2}v_0& \epsilon u^*_0\\
v^*_0&-\sqrt{2}u_0&v^*_0\end{array}\right)\]
is the gauge transform diagonalizing $S^{(0)}$, see \cite{yantih} for more explanations. The
dispersion law of a NLEE is an essential feature encoding the time dependence of its solutions and that
way it labels the NLEE within the integrable hierarchy. The dispersion law of (\ref{ghf}) is
\begin{equation}
f(\lambda) = - \lambda^2 \diag(1,-2,1)/3\,.
\end{equation}
The linear problems (\ref{bare1}) and (\ref{bare2}) will be referred to as bare (seed) linear problems 
and their fundamental solutions will be called bare (seed) fundamental solutions. We shall denote the set
of all bare fundamental solutions by $\mathcal{F}_0$.

As discussed in \cite{yantih}, conditions (\ref{red10}) and (\ref{red20}) are due to the action of the reduction
group $\bbz_2\times\bbz_2$ on the set of bare fundamental solutions, see \cite{mikh1,mikh2} for more explanations.
Indeed, in our case the following $\bbz_2\times\bbz_2$-action 
\begin{eqnarray}
\psi_0(x,t,\lambda) &\to& H\psi_0(x,t,-\lambda)H,\label{psi_0_red1}\\
\psi_0(x,t,\lambda) &\to& Q_{\epsilon}\hat{\psi}^{\dag}_0(x,t,\lambda^*)Q_{\epsilon}
\label{psi_0_red2}
\end{eqnarray}
where $\hat{X}$ stands for the inverse of a matrix $X$, leads to (\ref{red10}) and (\ref{red20})
respectively.

Now, let us apply the gauge (dressing) transform
\[\mathcal{G}:\mathcal{F}_0 \to \mathcal{F}_1 = \mathcal{G}\mathcal{F}_0
:= \{\mathcal{G}\psi_0| \psi_0\in\mathcal{F}_0\}\]
where $\mathcal{G}(x,t,\lambda)$ is a $3\times 3$-matrix with unit determinant. Since the Lax operators
are transformed through:
\begin{equation}
L_0 \to L_1 := \mathcal{G}L_0\hat{\mathcal{G}}\, ,\qquad A_0 \to A_1 := \mathcal{G}A_0\hat{\mathcal{G}}\, ,
\label{lax_dres}\end{equation}
the operators $L_1$ and $A_1$ commute as well. We shall require that the auxiliary linear problems
remain covariant under dressing transform, i.e. we have 
\begin{equation}
L_1(\lambda) \psi_1(x,t,\lambda) = 0\, , \qquad A_1(\lambda)\psi_1(x,t,\lambda) 
= \psi_1(x,t,\lambda)f(\lambda)\,,\qquad \psi_1\in\mathcal{F}_1
\label{dres}
\end{equation}
for $L_1$ and $A_1$ being of the form
\begin{eqnarray}
L_1(\lambda) & = & \rmi\partial_x - \lambda S^{(1)}, \qquad 
S^{(1)} = \left(\begin{array}{ccc}
0 & u_1 & v_1\\ 
\epsilon u^*_1 & 0 & 0 \\
v^*_1 & 0 & 0\end{array}\right), \label{lax1_dres}\\
A_1(\lambda) & = & \rmi\partial_t + \sum_{k=1}^{N}\lambda^k A^{(1)}_k. \label{lax2_dres}
\end{eqnarray} 
Above, $u_1$ and $v_1$ are some new (unknown) functions satisfying the same NLEE. This is the
main idea underlying the method to construct new solutions from a known one. Thus, dressing
method can symbolically be presented as the following sequence of steps:
\[(u_0,v_0)\to (L_0, A_0) \to\psi_0\stackrel{\mathcal{G}}{\to}\psi_1 \to (L_1, A_1)\to (u_1,v_1)\, .\]

Though the covariance of the linear problems is a strong condition, it does not determine the
dressing factor $\mathcal{G}$ uniquely. Indeed, after comparing (\ref{bare1}) and (\ref{bare2})
with (\ref{dres}), we see that $\mathcal{G}$ must solve the system of linear partial
differential equations: 
\begin{eqnarray}
\rmi\partial_x \mathcal{G} & - & \lambda \left(S^{(1)}\mathcal{G} - \mathcal{G}S^{(0)}\right) = 0\, ,
\label{g_pde1}\\
\rmi\partial_t \mathcal{G} & + & \sum_{k=1}^{N}\lambda^k \left( A^{(1)}_k \mathcal{G}
- \mathcal{G} A^{(0)}_k\right) = 0\,.
\label{g_pde2}
\end{eqnarray}
Equations (\ref{g_pde1}) and (\ref{g_pde2}) tell us nothing about the $\lambda$-dependence of
$\mathcal{G}$. This is why we need to make a few assumptions about its behavior with respect to
$\lambda$ in order to obtain more specific results. Let us assume the dressing factor and its
derivatives in $x$ and $t$ are defined in the vicinity of $\lambda = 0$. After setting
$\lambda=0$ in (\ref{g_pde1}) and (\ref{g_pde2}), we immediately see that
\[\partial_x\mathcal{G}|_{\lambda = 0} = \partial_t\mathcal{G}|_{\lambda = 0} = 0.\]
Those relations imply $\mathcal{G}$ should depend non-trivially on $\lambda$ otherwise it will be
merely a constant. We shall pick up $\mathcal{G}|_{\lambda = 0} = \openone$ since it does not lead
to any loss of generality. In fact, that normalization corresponds to the normalization of fundamental
analytic solutions at $\lambda = 0$, see \cite{yantih}.

Equation (\ref{g_pde1}) allows one to find $u_1$ and $v_1$ in terms of the seed solution and the dressing
factor. At this point, we require that $\mathcal{G}$ as well as its derivatives in $x$ and $t$ are regular
when $|\lambda|\to \infty$. After dividing (\ref{g_pde1}) by $\lambda$ and setting $|\lambda|\to\infty$ we
derive the following interrelation:
\begin{equation}
S^{(1)} = \mathcal{G}_{\infty}S^{(0)}\hat{\mathcal{G}}_{\infty},\qquad
\mathcal{G}_{\infty}(x,t) := \lim_{|\lambda|\to\infty}\mathcal{G}(x,t,\lambda)\, .
\label{s10_eq}\end{equation}
Since $S^{(0)}$ is determined by $(u_0,v_0)$ and $S^{(1)}$ is determined by $(u_1,v_1)$, the above
relation allows us to construct another solution of our system starting from a known one. 

The form of the dressed operators and the zero curvature condition imply $L_1$ and $A_1$ obey the 
symmetries (\ref{red10}) and (\ref{red20}) too. Therefore, the set of dressed fundamental
solutions is subject to the $\bbz_2\times\bbz_2$-action 
\begin{eqnarray}
\psi_1(x,t,\lambda) &\to& H\psi_1(x,t,-\lambda)H,\qquad \psi_1\in\mathcal{F}_1\, ,\label{psi_1_red1}\\
\psi_1(x,t,\lambda) &\to& Q_{\epsilon}\hat{\psi}^{\dag}_1(x,t,\lambda^*)Q_{\epsilon}\, .
\label{psi_1_red2}
\end{eqnarray}
The $\bbz_2\times\bbz_2$ action on $\mathcal{F}_0$ and $\mathcal{F}_1$ implies the dressing
factor is not entirely arbitrary but obeys the symmetry relations:
\begin{eqnarray}
H\mathcal{G}(x,t,-\lambda)H &=& \mathcal{G}(x,t,\lambda)\,,\label{g_red1}\\
Q_{\epsilon}\mathcal{G}^{\dag}(x,t,\lambda^*)Q_{\epsilon} &=& \hat{\mathcal{G}}(x,t,\lambda)\, .
\label{g_red2}\end{eqnarray}

A simple ansatz for dressing factor meeting all the requirements discussed so far is given by:
\begin{equation}
\mathcal{G}(x,t,\lambda) = \openone + \lambda\sum_{j}\left[ \frac{ B_j(x,t)}{\mu_j(\lambda - \mu_j)}
+ \frac{HB_j(x,t)H}{\mu_j(\lambda + \mu_j)}\right],
\qquad \mu_j\in\bbc\backslash\{0\}\, .
\label{dress_fac}
\end{equation}

Generally speaking, dressing procedure {\bf does not} preserve the spectrum of scattering operator, see
\cite{book}. Indeed, let us assume for simplicity that $L_0$ has no discrete eigenvalues, i.e. the spectrum
of the bare scattering operator coincides with the real line, see \cite{yantih}. Denote the resolvent of
$L_0$ by $R_0(\lambda)$. According to (\ref{lax_dres}), dressing transform maps the bare resolvent onto  
\begin{equation}
R_1(\lambda) = \mathcal{G}R_0(\lambda)\hat{\mathcal{G}}\, .
\label{resolv_dres}\end{equation}
Equation (\ref{resolv_dres}) implies the singularities of the dressing factor and its inverse "produce" singularities
of the dressed resolvent operator $R_1(\lambda)$. This way dressing procedure adds new discrete eigenvalues
to the bare scattering operator. 

In order to find $L_1 $ and $A_1$ through (\ref{s10_eq}), we need to know $B_j(x,t)=\Res(\mathcal{G}(x,t,\lambda);\mu_j)$.
The algorithm to find the residues of (\ref{dress_fac}) consists in two steps. In the first step, one considers the identity $\mathcal{G}\hat{\mathcal{G}}=\openone$ which gives rise to a set of algebraic relations for $B_j$. The form of these relations
crucially depends on the location of $\mu_j$ --- whether the poles of the dressing factor are arbitrary complex numbers
with nonzero real and imaginary parts (generic case) or the poles are either imaginary or real numbers. If the poles are
complex numbers in generic position or purely imaginary then we obtain soliton-like solutions. For poles lying on the real
line, we have degeneracy in the spectrum of the scattering operator. As a result, we obtain quasi-rational solutions. Due to
all these differences, we shall consider the three cases in separate subsections. 

Though the algebraic relations may be different, they always imply that the residues of $\mathcal{G}$ are some singular
(degenerate) matrices which could be decomposed as follows:
\begin{equation}
B_j(x,t) = X_j(x,t)F^T_j(x,t)
\label{res_decomp}\end{equation}
where $X_j(x,t)$ and $F_j(x,t)$ are rectangular matrices of certain rank and the superscript $T$ stands for matrix transposition.
After substituting (\ref{res_decomp}) into the algebraic relations, we are able to express $X_j$ through $F_j$.
 
The factors $F_j$ are determined in the second step. For this to be done, we consider differential equations
(\ref{g_pde1}) and (\ref{g_pde2}). Like for the algebraic relations discussed above, the calculations here depend on the
location of the poles of $\mathcal{G}$. However, equation (\ref{g_pde1}) always leads to the following result 
\begin{equation}
F^T_j(x,t) = F^T_{j,0}(t)\hat{\psi_0}(x,t,\mu_j)
\end{equation}
allowing one to construct $F_j$ through a bare fundamental solution defined in the vicinity of $\mu_j$ and some arbitrary $x$-independent matrices $F_{j,0}$. The matrices $F_{j,0}$ depend on $t$ in a way governed by equation (\ref{g_pde2}). Regardless of the location of $\mu_j$, we derive exponential $t$-dependence. Thus, to recover the time dependence in all formulas we may use the rule below:
\begin{equation}
F^T_{j,0}(t)\quad\to\quad F^T_{j,0}(t)\rme^{-\rmi f(\mu_j)t}
\end{equation}
where $f(\lambda)$ is the dispersion law of the NLEE, see (\ref{disp_law}). We shall demonstrate that
procedure in the subsections to follow.

\subsection{Soliton Type Solutions I. Generic Case}\label{sol_gen}

Let us start with the case when the poles of (\ref{dress_fac}) are complex numbers in generic position,
i.e. $\mu^2_j\notin\bbr$ for all $j$. From the symmetry condition (\ref{g_red2}), we immediately deduce
that
\begin{equation}
\hat{\mathcal{G}}(x,t,\lambda) = \openone + \lambda\sum_i\left[\frac{Q_{\epsilon}B^{\dag}_i(x,t)Q_{\epsilon}}
{\mu^*_i(\lambda - \mu^*_i)} + 	\frac{Q_{\epsilon}HB^{\dag}_i(x,t)HQ_{\epsilon}}
{\mu^*_i(\lambda + \mu^*_i)}\right].
\label{g_inv1}\end{equation}
Thus, the dressing factor and its inverse have poles located at different points. 

Let us consider the identity $\mathcal{G}(x,t,\lambda)\hat{\mathcal{G}}(x,t,\lambda)=\openone$. Since it holds identically
in $\lambda$, the residues of $\mathcal{G}\hat{\mathcal{G}}$ should vanish. After evaluating the residue of
$\mathcal{G}\hat{\mathcal{G}}$ at $\mu^*_i$ we easily obtain the following algebraic relation:
\begin{equation}
\left[\openone + \mu^*_i\sum_j\left(\frac{B_j}{\mu_j(\mu^*_i - \mu_j)}
+ \frac{HB_jH}{\mu_j(\mu^*_i + \mu_j)}\right)\right]Q_{\epsilon}B^{\dag}_iQ_{\epsilon} = 0\, .	
\label{algsys1}\end{equation}
Evaluation of the residues at $\pm\mu_i$ and $-\mu_i^*$ leads to equations that can easily be reduced
to (\ref{algsys1}), thus giving us no new constraints.

It is seen from (\ref{algsys1}) that each $B_i$ should be a degenerate matrix, hence it can be factored 
\begin{equation}
B_i(x,t)=X_i(x,t)F^T_i(x,t)
\label{b_fac}\end{equation}
where $X_i$ and $F_i$ are two rectangular matrices. After substituting (\ref{b_fac}) into (\ref{algsys1}), one obtains the
linear system 
\begin{equation}
Q_{\epsilon}F^*_i = \mu^*_i\sum_j\left(X_j\frac{F^T_jQ_{\epsilon}F^*_i}{\mu_j(\mu_j - \mu^*_i)}
- HX_j\frac{F^T_jHQ_{\epsilon}F^*_i}{\mu_j(\mu^*_i + \mu_j)}\right)
\label{matr_sys1}\end{equation}
for the factors $X_j$. Solving it, allows one to express $X_j$ through $F_j$. This is easier when the dressing factor has
just a single pair of poles: $\mu$ and $-\mu$. Assuming $X$ and $F$ are column-vectors, we get:
\begin{equation}
X =	\left(\frac{\mu^*F^TQ_{\epsilon}F^*}{\mu(\mu - \mu^*)}
 - \frac{\mu^*F^T HQ_{\epsilon}F^*}{\mu(\mu + \mu^*)}H\right)^{-1}Q_{\epsilon}F^* . 
\label{XF_sys}\end{equation}

Let us return now to the general case. The factors $F_j$ can be found from differential equation
(\ref{g_pde1}). For that purpose we rewrite (\ref{g_pde1}) in the following way
\begin{equation}
\lambda S^{(1)}= \rmi\partial_x \mathcal{G}\hat{\mathcal{G}} + \lambda\mathcal{G}S^{(0)}\hat{\mathcal{G}}	
\end{equation}
and calculate the residues of its right-hand side at $\lambda=\mu_i$. After taking into account
(\ref{matr_sys1}), we obtain the linear differential equation
\begin{equation}
\rmi\partial_x F^T_i + \mu_i F^T_iS^{(0)} = 0 
\label{F_res}\end{equation}
which is immediately solved to give
\begin{equation}
F^T_i(x,t) = F^T_{i, 0}(t)\hat{\psi}_0(x,t,\lambda=\mu_i)\, .
\label{f_psi0}\end{equation}
Above, $F_{i,0}$ are "integration constant" matrices which depend on $t$ however. Evaluation of the residues at $-\mu_i$
or $\pm\mu^*_i$ gives equations that are easily reduced to (\ref{F_res}) after taking into account the symmetries of $S^{(0)}_1$.

The $t$-dependence of $F_{i,0}$ is determined from equation (\ref{g_pde2}) after it is rewritten as follows:
\begin{equation}
\sum_{k=1}^{N}\lambda^k A^{(1)}_k = - \rmi\partial_t \mathcal{G}\hat{\mathcal{G}}
+ \mathcal{G} \sum_{k=1}^{N}\lambda^k A^{(0)}_k \hat{\mathcal{G}}\, .
\label{g_pde2_2}
\end{equation}
Evaluation of the residues of both hand sides of (\ref{g_pde2_2}) for $\lambda=\mu_i$ gives 
\begin{equation}
\rmi\partial_tF^T_i - F^T_i\sum^N_{k=1}\mu^k_i A^{(0)}_k = 0\, .
\label{fj_t}\end{equation}
Yet again we do not need to consider the residues for $\lambda = \pm\mu^*_i$ and $\lambda = -\mu_i$ due to the symmetries
of the coefficients of $A(\lambda)$.

After substituting (\ref{f_psi0}) into (\ref{fj_t}), we derive a linear differential equation for $F_{i,0}$, namely
\begin{equation}
\rmi\partial_t F^T_{i,0} - F^T_{i,0} f(\mu_i) = 0
\label{fj0_t}
\end{equation}
where $f(\lambda)$ is the dispersion law of the NLEE. Equation (\ref{fj0_t}) is easily solved and allows us to state the
rule: in order to recover the $t$-dependence in all formulas, one has to make the following substitution
\begin{equation}
F^T_{i,0}\quad\to\quad F^T_{i,0}\rme^{-\rmi f(\mu_i)t} .
\label{f_j0_evol}\end{equation}

Now, let us apply the general results obtained above to some seed solution. An obvious choice for
seed solution satisfying (\ref{constr}) and (\ref{triv_back}) is
\begin{equation}
S^{(0)} = \left(\begin{array}{ccc}
0 & 0 & 1 \\ 0 & 0 & 0 \\
1 & 0 & 0
\end{array}\right).
\label{seed}
\end{equation}
One can prove that the bare scattering operator in this case has no discrete eigenvalues.

For a bare fundamental solution we shall choose
\begin{equation}
\psi_0(x,\lambda) = \left(\begin{array}{ccc}
\cos{\lambda x} & 0 & - \rmi\sin{\lambda x} \\ 0 & 1 & 0 \\
- \rmi\sin{\lambda x} & 0 & \cos{\lambda x}
\end{array}\right).	
\label{psi_0}\end{equation}
That fundamental solution is invariant with respect to both reductions (\ref{psi_0_red1})
and (\ref{psi_0_red2}) which makes it rather convenient for the calculations to follow.

From that point on, we shall focus on the simplest case when the dressing factor has a single pair of
poles and $X$ and $F$ are column-vectors. After substituting (\ref{XF_sys}) and (\ref{b_fac}) into
(\ref{dress_fac}) and (\ref{g_inv1}), equation (\ref{s10_eq}) can be written in components as follows:
\begin{eqnarray}
u_1 &=& - \frac{\left[\mu |F^1|^2 + \mu^*(\epsilon|F^2|^2 + |F^3|^2)\right]\left(\mu^2-(\mu^*)^2\right)F^2\left(F^3\right)^*}{\mu^*\left[\mu^*|F^1|^2 + \mu (\epsilon |F^2|^2 + |F^3|^2)\right]^2}\, ,\label{u1_gen}\\
v_1 &=& \frac{\left[\mu |F^1|^2 + \mu^*(\epsilon|F^2|^2 + |F^3|^2)\right]}{\mu^*\left[\mu^*|F^1|^2 + \mu (\epsilon |F^2|^2 + |F^3|^2)\right]^2}\times\nonumber\\
&&\left\{\mu^*\left[\mu|F^1|^2 + \mu^*(\epsilon|F^2|^2 + |F^3|^2)\right] + \left(\mu^2 - (\mu^*)^2\right)\epsilon|F^2|^2\right\}\label{v1_gen}
\end{eqnarray}
where $F^T = (F^1, F^2, F^3)$. It is not hard to check that (\ref{u1_gen}) and (\ref{v1_gen}) satisfy the constraint (\ref{constr}) for any $F$.

On the other hand, (\ref{f_psi0}) leads to the following expression for $F$:  
\begin{equation}
F(x) = \left(\begin{array}{c}
F_{0}^{1}\cos\mu x + \rmi F_{0}^{3}\sin \mu x \\
F_{0}^{2} \\ F_{0}^{3}\cos \mu x + \rmi F_{0}^{1}\sin \mu x 
\end{array}\right),\qquad F_0	= \left(\begin{array}{c}
F_{0}^{1} \\ F_{0}^{2} \\ F_{0}^{3} \end{array}\right).
\label{f_sol}
\end{equation}
In order to evaluate $X$, we shall need the following quadratic expressions as well:
\begin{eqnarray}
F^TQ_{\epsilon}F^* & = & \left(\left |F_{0}^{1}\right |^2 + |F_{0}^{3}|^2\right)\cosh{2\kappa x}
- 2|F_{0}^{1}F_{0}^{3}|\cos{\varphi}\sinh{2\kappa x} + \epsilon |F_{0}^{2}|^2,\label{ff1}\\
F^TQ_{\epsilon}HF^* & = & \left[\left(|F_{0}^{3}|^2 - |F_{0}^{1}|^2\right)\cos{2\omega x} 
- 2|F_{0}^{1}F_{0}^{3}|\sin{\varphi}\sin{2\omega x} + \epsilon|F_{0}^{2}|^2 \right]\label{ff2}
\end{eqnarray}
where $\omega=\re\mu>0$, $\kappa=\im\mu>0$ and $\varphi = \arg F_{0}^{1} - \arg F_{0}^{3}$.  Taking into account
the structure of (\ref{psi_0}), (\ref{f_sol}) and (\ref{u1_gen}), (\ref{v1_gen}), it is natural to consider in
detail the following three elementary cases, see \cite{side9}.  
\begin{enumerate}
\item First, let us assume $F_{0}^{2} = 0$ and $F_{0}^{1}\neq \pm F_{0}^{3}$. Then for $F$ we have
\begin{equation}
F(x) = \left(\begin{array}{c}
F_{0}^{1}\cos\mu x  + \rmi F_{0}^{3}\sin \mu x \\
0 \\ F_{0}^{3}\cos \mu x + \rmi F_{0}^{1}\sin \mu x 
\end{array}\right)
\label{f_sol1}\end{equation}
and expressions \eqref{ff1} and \eqref{ff2} now could be simplified to
\begin{eqnarray}
F^TQ_{\epsilon}F^* & = & A\cosh(2\kappa x + \xi_0)\, ,\label{ff1a}\\
F^TQ_{\epsilon}HF^* & = & - A\cos(2\omega x + \delta_0)\label{ff2a}
\end{eqnarray}
where
\begin{eqnarray}
\cosh{\xi_0} &=& \frac{|F_{0}^{1}|^2 + |F_{0}^{3}|^2}{A}\; ,
\qquad\sinh{\xi_0} = - \frac{2|F_{0}^{1}F_{0}^{3}|\cos\varphi}{A}\; ,\label{xi_0}\\
A &=& \sqrt{\left(|F_{0}^{1}|^2 + |F_{0}^{3}|^2\right)^2 - 4|F_{0}^{1}F_{0}^{3}|^2\cos^2\varphi}
\; , \label{a_def}\\
\cos{\delta_0} &=& \frac{|F_{0}^{1}|^2 - |F_{0}^{3}|^2}{A}\; ,
\qquad\sin{\delta_0} = - \frac{2|F_{0}^{1}F_{0}^{3}|\sin\varphi}{A}\;.\label{delta_0}
\end{eqnarray}
After substituting (\ref{f_sol1})--(\ref{delta_0}) into (\ref{XF_sys}) and  (\ref{s10_eq}), we get the dressed solution
at a fixed moment of time:
\begin{eqnarray}
u_1(x) & = & 0\, , \label{u1_1}\\
v_1(x) & = & \exp\left[4\rmi\arctan\frac{\kappa\cos(2\omega x + \delta_0)}
{\omega\cosh(2\kappa x + \xi_0)}\right] .
\label{v1_1}\end{eqnarray}
To recover the time evolution in (\ref{u1_1}) and (\ref{v1_1}), one needs to make the substitution:
\begin{equation}
\varphi	\to \varphi,\qquad \xi_0\to\xi_0,\qquad \delta_0\to\delta_0,\qquad
A\to A\exp\left(\frac{-4\omega\kappa t}{3}\right)
\label{par_evol1}\end{equation}
that follows directly from (\ref{f_j0_evol}). It is seen that (\ref{u1_1}) and (\ref{v1_1}) remain invariant under
transformation (\ref{par_evol1}) so the dressed solution is stationary and non-singular. The solution just derived
coincides with that found in \cite{side9} (see formula (3.26)) for the generalized HF with Hermitian reduction. 

\item Let us assume $F_{0}^{2} \neq 0$ . Without any loss of generality we could just set
$F_{0}^{2} = 1$. There exist two elementary options: $F_{0}^{1} = F_{0}^{3}$ and $F_{0}^{1} = - F_{0}^{3}$ . Let us
consider the former one. After recovering the time evolution, the vector $F(x)$ is given by:
\begin{equation}
F(x,t) = \left(\begin{array}{c}
F_{0}^{1} \rme^{\rmi\mu x + \frac{\rmi \mu^2 t}{3}}\\
\rme^{-\frac{2\rmi \mu^2 t}{3}} \\ F_{0}^{1}\rme^{\rmi\mu x + \frac{\rmi \mu^2 t}{3}}
\end{array}\right).
\end{equation}
The solution in this case reads:
\begin{eqnarray}
u_1(x,t) &=& \frac{4\omega\kappa \left[2\omega\rme^{\vartheta(x,t)}
+ \epsilon(\omega - \rmi\kappa)\rme^{- \vartheta(x,t)}\right]
\rme^{-\rmi[\omega x + (\omega^2 - \kappa^2)t + \delta]}}
{(\omega - \rmi\kappa)\left[2\omega\rme^{\vartheta(x,t)}
+ \epsilon(\omega + \rmi\kappa)\rme^{- \vartheta(x,t)}\right]^2} \; ,
\label{u1_2}\\
v_1(x,t) & = & 1 - \frac{8\epsilon\omega\kappa^2}{(\omega - \rmi\kappa)
\left[2\omega\rme^{\vartheta(x,t)} + \epsilon(\omega + \rmi\kappa)\rme^{-\vartheta(x,t)}\right]^2}
\label{v1_2}\end{eqnarray}
where
\[\vartheta(x,t) = -\kappa(x + 2\omega t) + \ln |F_{0,1}|\, ,\qquad
\delta = \left(\frac{\pi}{2} + \arg{F_{0,1}}\right).\]
Solution (\ref{u1_2}), (\ref{v1_2}) has no singularities and goes into the one obtained in \cite{side9} (see equation (3.27))
when $\epsilon = 1$. 

\item  Assume now $F_{0}^{2} = 1$  and $F_{0}^{1} = - F_{0}^{3}$ . 

In this case $F(x,t)$ is given by:
\begin{equation}
F(x,t) = \left(\begin{array}{c}
F_{0}^{1} \rme^{-\rmi\mu x + \frac{\rmi \mu^2 t}{3}}\\
\rme^{-\frac{2\rmi \mu^2 t}{3}} \\ - F_{0}^{1}\rme^{-\rmi\mu x + \frac{\rmi \mu^2 t}{3}}
\end{array}\right).
\end{equation}
The corresponding dressed solution looks as follows:
\begin{eqnarray}
u_1(x,t) &=& \frac{4\omega\kappa \left[2\omega\rme^{\tilde{\vartheta}(x,t)}
+ \epsilon(\omega - \rmi\kappa)\rme^{- \tilde{\vartheta}(x,t)}\right]
\rme^{\rmi[\omega x + (\kappa^2 - \omega^2)t + \tilde{\delta}]}}
{(\omega - \rmi\kappa)\left[2\omega\rme^{\tilde{\vartheta}(x,t)}
+ \epsilon(\omega + \rmi\kappa)\rme^{- \tilde{\vartheta}(x,t)}\right]^2} \; ,
\label{u1_3}\\
v_1(x,t) & = & 1 - \frac{8\epsilon\omega\kappa^2}{(\omega - \rmi\kappa)
\left[2\omega\rme^{\tilde{\vartheta}(x,t)} + \epsilon(\omega + \rmi\kappa)
\rme^{-\tilde{\vartheta}(x,t)}\right]^2}
\label{v1_3}\end{eqnarray}
where
\[\tilde{\vartheta}(x,t) = \kappa(x - 2\omega t) + \ln |F_{0,1}|\, ,\qquad
\tilde{\delta} = \frac{\pi}{2} - \arg F_{0,1}\, .\]
Formally, this solution could be derived from \eqref{u1_2}, \eqref{v1_2} by applying the transform
$x\to -x$ and $F_{0,1}\to F_{0,3}$ .  
\end{enumerate}

All the solutions constructed explicitly in this subsection are connected with four distinct poles (quadruples) of
$\mathcal{G}$ and $\hat{\mathcal{G}}$, i.e. four discrete eigenvalues of the dressed scattering operator. This is why
we call such solutions "quadruplet" solutions.

\subsection{Soliton Type Solutions II. Imaginary Poles}\label{doublet}

Let us assume now that all the poles of $\mathcal{G}$ are purely imaginary, i.e. we have
$\mu_j = \rmi \kappa_j$ for some real numbers $\kappa_j\neq 0$. Thus, we could write 
\begin{equation}
\mathcal{G}(x,t,\lambda) = \openone + \lambda\sum_{i}\left[\frac{B_i(x,t)}{\rmi\kappa_i(\lambda - \rmi\kappa_i)} + \frac{HB_i(x,t)H}{\rmi\kappa_i(\lambda + \rmi\kappa_i)}\right] \; .
\label{g_im}\end{equation}
for the dressing factor and
\begin{equation}
\hat{\mathcal{G}}(x,t,\lambda) = \openone - \lambda\sum_i\left[\frac{Q_{\epsilon}B^{\dag}_i(x,t)Q_{\epsilon}}
{\rmi\kappa_i(\lambda + \rmi\kappa_i)} + \frac{Q_{\epsilon}HB^{\dag}_i(x,t)HQ_{\epsilon}}
{\rmi\kappa_i(\lambda - \rmi\kappa_i)}\right]
\label{ginv_im}\end{equation}
for its inverse. Since $\mathcal{G}$ and $\hat{\mathcal{G}}$ have the same poles, the identity
$\mathcal{G}\hat{\mathcal{G}} = \openone$ gives rise to algebraic relations that do not follow from (\ref{algsys1}).
It is easily checked that the following equations hold true:
\begin{eqnarray}
\lim_{\lambda\to \rmi\kappa_i}(\lambda - \rmi\kappa_i)^2\left(\mathcal{G}\hat{\mathcal{G}} = \openone\right)
\quad&\Rightarrow&\quad B_iQ_{\epsilon}HB_i^{\dag} = 0\; , \label{algrel1} \\
\lim_{\lambda\to \rmi\kappa_i}\partial_{\lambda}(\lambda - \rmi\kappa_i)^2\left(\mathcal{G}\hat{\mathcal{G}}
= \openone\right)\quad&\Rightarrow&\quad B_iQ_{\epsilon}H\Omega_i^{\dag} - \Omega_iQ_{\epsilon}HB^{\dag}_i = 0
\label{algrel2}
\end{eqnarray} 
where
\[\Omega_i = \openone + \frac{B_i}{\rmi\kappa_i} + \sum_{j\neq i} \frac{\kappa_iB_j}{\rmi\kappa_j(\kappa_i - \kappa_j)}
+ \sum_j \frac{\kappa_iHB_jH}{\rmi\kappa_j(\kappa_i + \kappa_j)}\; .\]
There is no need to perform similar calculations for $\lambda = -\rmi\kappa_i$ since
all algebraic equations can be reduced to (\ref{algrel1}) and (\ref{algrel2}).

Relation (\ref{algrel1}) implies that $B_i(x,t)$ is degenerate so decomposition (\ref{b_fac}) holds again
for some rectangular matrices $X_i$ and $F_i$ (we shall assume that these matrices are simply column vectors).
Due to (\ref{algrel1}) the factors $F_i$ satisfy the quadratic relation
\begin{equation}
F^T_iQ_{\epsilon}HF^*_i = 0\, .
\label{f_sq}\end{equation}
After substituting (\ref{b_fac}) into (\ref{algrel2}), we see that there exist functions $\alpha_i$ such that:
\begin{equation}
\Omega_iQ_{\epsilon}HF^*_i = -X_i\alpha_i\, ,\qquad
\alpha_i^* = \alpha_i \, .	
\label{algrel2a}
\end{equation}
That relation allows one to find all vectors $X_i$ in terms of $F_i$ and 
$\alpha_i$. In the simplest case when the dressing factor has a single pair of
poles $\pm\rmi\kappa$ (\ref{algrel2a}) is easily solvable to give:
\begin{equation}
X = - \left(\alpha + \frac{F^TQ_{\epsilon}F^*}{2\rmi\kappa} H\right)^{-1}Q_{\epsilon}HF^* \; .	
\label{fx_sys2}
\end{equation}
After taking into account (\ref{g_im}), (\ref{ginv_im}), (\ref{b_fac}) and (\ref{fx_sys2}), equation
(\ref{s10_eq}) could be written down in components as follows:
\begin{eqnarray}
u_1(x,t) &=& \frac{2(|F^1(x,t)|^2 + \rmi\kappa\alpha(x,t))F^2(x,t)\left(F^3(x,t)\right)^*}{\left(|F^1(x,t)|^2
	 - \rmi\kappa\alpha(x,t)\right)^2}	\;,\label{u1_doubl}\\
v_1(x,t) &=& 1 + \frac{2\left(2\rmi\kappa\alpha(x,t) - \epsilon |F^2(x,t)|^2\right)}{|F^1(x,t)|^2
	 - \rmi\kappa\alpha(x,t)}
+ \frac{4\rmi\kappa\alpha(x,t)\left(\rmi\kappa\alpha(x,t) - \epsilon |F^2(x,t)|^2\right)}{\left(|F^1(x,t)|^2
	 - \rmi\kappa\alpha(x,t)\right)^2} . \label{v1_doubl}
\end{eqnarray}
It is easy to check that (\ref{u1_doubl}) and (\ref{v1_doubl}) are interrelated through
$|v|^2 + \epsilon |u|^2 = 1$ for any real-valued $\alpha$ and $F$ obeying condition (\ref{f_sq}).

Like in the generic case (see previous subsection) we can find $F_i$ and $\alpha_i$ by analyzing (\ref{g_pde1})
and (\ref{g_pde2}). Considerations similar to those in the generic case lead to the following linear differential equations:
\begin{eqnarray}
\lim_{\lambda\to \rmi\kappa_i}(\lambda - \rmi\kappa_i)^2\left( \rmi\partial_x \mathcal{G}\hat{\mathcal{G}} + \lambda\mathcal{G}S^{(0)}\hat{\mathcal{G}} = \lambda S^{(1)}\right)\quad&\Rightarrow&\quad \rmi\partial_x F^T_i + \mu_i F^T_iS^{(0)} = 0 
\, , \label{difrel1} \\
\lim_{\lambda\to \rmi\kappa_i}\partial_{\lambda}(\lambda - \rmi\kappa_i)^2\left(\rmi\partial_x \mathcal{G}\hat{\mathcal{G}} + \lambda\mathcal{G}S^{(0)}\hat{\mathcal{G}} = \lambda S^{(1)}\right)\quad&\Rightarrow&\quad
\rmi\partial_x\alpha_i = F^T_iS^{(0)}Q_{\epsilon}HF^*_i \,. 	
 \label{difrel2}
\end{eqnarray}
Equation (\ref{difrel1}) shows that (\ref{f_psi0}) holds true for the vectors $F_i$ while (\ref{difrel2}) implies that
the scalar functions $\alpha_i$ can be expressed in terms of the seed fundamental solution as given by:
\begin{equation}
\alpha_i(x) = \alpha_{i,0} + F^T_{i,0}\hat{\psi}_0(x,\rmi\kappa_i)\frac{\partial\psi_0(x,\rmi\kappa_i)}{\partial\lambda}K_{\psi_0}
Q_{\epsilon}HF^*_{i,0}\; .	
\label{alphai_res}\end{equation}
Above, $\alpha_{i,0}$ is an integration constant and $K_{\psi}(\lambda)
= \hat{\psi}(\lambda)HQ_{\epsilon}\hat{\psi}^{\dag}(-\lambda^*)Q_{\epsilon}H$ measures the "deviation" of the solution $\psi$ from
invariant solutions, i.e. if $\psi$ is invariant with respect to reductions \eqref{psi_0_red1} and
\eqref{psi_0_red2} then $K_{\psi} = \openone$.

Finally, we need to recover the time dependence in all formulas. For this to be done we consider equation
(\ref{g_pde2}) as we did in previous subsection. As a result, we derive a set of equations for $F_i$ and
$\alpha_i$, namely:
\begin{eqnarray}
&&\lim_{\lambda\to\rmi\kappa_i}(\lambda - \rmi\kappa_i)^2\left[-\rmi\partial_t\mathcal{G}\hat{\mathcal{G}}
 + \mathcal{G}\sum_k \lambda^k A^{(0)}_k \hat{\mathcal{G}}\right] =0\qquad\Rightarrow\quad\nonumber\\
&&\rmi\partial_tF^T_i - F^T_j\sum_k(\rmi\kappa_i)^k A^{(0)}_k = 0\, ,\label{fj_t2}\\
&&\lim_{\lambda\to\rmi\kappa_i}\partial_{\lambda}\left\{(\lambda - \rmi\kappa_i)^2\left[-\rmi\partial_t\mathcal{G}\hat{\mathcal{G}}
+ \mathcal{G}\sum_k\lambda^k A^{(0)}_k\hat{\mathcal{G}}\right]\right\}=0\qquad\Rightarrow\quad\nonumber\\
&&\rmi\partial_t\alpha_i + F^T_i\sum_k k(\rmi\kappa_i)^{k-1}A^{(0)}_k Q_{\epsilon} HF^*_i = 0\, .
\label{alphaj_t}\end{eqnarray}
Equation (\ref{fj_t2}) coincides with (\ref{fj_t}), thus, after substituting (\ref{f_psi0}) into
(\ref{fj_t2}), we derive (\ref{fj0_t}) and rule (\ref{f_j0_evol}) holds again. On the other hand,
(\ref{alphaj_t}) is reduced to
\begin{equation}
\rmi\partial_t\alpha_{i,0} + F^T_{i,0}\frac{\rmd f(\rmi\kappa_i)}{\rmd \lambda}K_{\psi_0}HQ_{\epsilon}F^*_{i,0} = 0\, .	
\label{alphaj0_t}\end{equation}
If bare fundamental solution is invariant with respect to both reductions, i.e. $K_{\psi_0} = \openone$, then
(\ref{alphaj0_t}) simplifies to
\begin{equation}
\rmi\partial_t\alpha_{i,0} + F^T_{i,0}\frac{\rmd f(\rmi\kappa_i)}{\rmd \lambda}HQ_{\epsilon}F^*_{i,0} = 0 \, .
\label{alphaj0_t2}	
\end{equation}
 
Let us consider the case when the seed solution $S^{(0)}$ is picked up in the form (\ref{seed}) and the corresponding fundamental
solution is given by (\ref{psi_0}). As we discussed in previous subsection, for (\ref{psi_0}) we have $K_{\psi_0} = \openone$.
From this point on, we shall assume that $\mathcal{G}$ has a single pair of imaginary poles. In this case the 3-vector $F$ reads:
\begin{equation}
F(x) = \left(\begin{array}{c}
\cosh\kappa x F_{0}^{1} - \sinh\kappa x F_{0}^{3} \\ F_{0}^{2} \\
\cosh\kappa x F_{0}^{3} - \sinh\kappa x F_{0}^{1}
\end{array}\right)	,\qquad F_0 = \left(\begin{array}{c}
F_{0}^{1} \\ F_{0}^{2} \\ F_{0}^{3} 
\end{array}\right) .
\label{fpsi_0_2}\end{equation}
Our further considerations depend on whether or not $F_{0}^{2}$ is equal to $0$. 

\begin{enumerate}
	
\item Let us first assume $F_{0}^{2} = 0$\,. Then we may set $|F_{0}^{1}| = |F_{0}^{3}| = 1$ and
equations \eqref{fpsi_0_2} and \eqref{alphai_res} now give
\begin{eqnarray}
|F^1(x)|^2 &=& \cosh 2\kappa x - \sinh 2\kappa x \cos2\varphi\, ,\quad 2\varphi = \arg F_{0}^{1}
- \arg F_{0}^{3} \, ,
\label{fmod_doubl1}\\
\alpha(x) &=& \alpha_0 + 2x\sin2\varphi	
\label{alpha_doubl1}
\end{eqnarray}
where $\alpha_0$ is $t$-independent in this case. After substituting (\ref{fmod_doubl1})
and (\ref{alpha_doubl1}) into (\ref{u1_doubl}) and (\ref{v1_doubl}), then taking into account (\ref{fpsi_0_2})
we get the following stationary solution 
\begin{eqnarray}
u_1(x) &=& 0 \, ,\label{u1_doubl2}\\
v_1(x) &=& \left[\frac{\cosh 2\kappa x - \sinh 2\kappa x \cos2\varphi
+ \rmi\kappa(\alpha_0 + 2 x\sin2\varphi)}{\cosh 2\kappa x - \sinh 2\kappa x \cos2\varphi
- \rmi\kappa(\alpha_0 + 2 x\sin2\varphi)}\right]^2 \; .
\label{v1_doubl2}	
\end{eqnarray}
The dressed solution just derived does not "feel" the presence of $\epsilon$, i.e. whether the reduction is Hermitian
or pseudo-Hermitian, hence it formally coincides with the one obtained in \cite{side9}. The right hand-side of
(\ref{v1_doubl2}) could be rewritten as:
\[v_1(x) = \exp\left[4 \rmi\arctan\left(\frac{\kappa(\alpha_0 + 2 x\sin2\varphi)}{\cosh 2\kappa x
- \sinh 2\kappa x \cos2\varphi}\right) \right].\]
When we have $F_{0,1} = \pm F_{0,3}$ , it simplifies to
\begin{equation}
v_1(x) =  \left[\frac{\rme^{\mp 2\kappa x} + \rmi\kappa\alpha_0}{\rme^{\mp 2\kappa x}
- \rmi\kappa\alpha_0 }\right]^2.
\end{equation}
It is seen that $v_1$ is nontrivial only if $\alpha_0\neq 0$ .

\item Assume now that $F_{0}^{2}\neq 0$\,. Thus, we may simply set $F_{0}^{2} = 1$\,.

Our further analysis depends on the value of $\epsilon$. Let us pick up $\epsilon = 1$ first. Due to
(\ref{f_sq}) and the invariance of (\ref{psi_0}) under the reductions, we have
\begin{equation}
F^T_0Q_{\epsilon}HF^*_0 = 0 \qquad\Rightarrow\qquad |F_{0}^{1}|^2 - |F_{0}^{3}|^2 = 1 \, .
\label{f0sq}\end{equation}
A natural parametrization for $F_{0}^{1}$ and $F_{0}^{3}$ is 
\begin{equation}
F_{0}^{1} = \cosh\theta \rme^{\rmi(\delta + \varphi)}, \qquad F_{0}^{3} = \sinh\theta\rme^{\rmi(\delta - \varphi)},
\qquad \theta > 0.         
\label{doubl_par}\end{equation}
Then, using (\ref{doubl_par}) we obtain
\begin{eqnarray}
|F^1(x,t)|^2 &=& \cosh^2(\kappa x - \theta)\cos^2\varphi + \cosh^2(\kappa x + \theta)\sin^2\varphi \, ,\\
\alpha(x,t) &=&  x\sinh2\theta\sin2\varphi - 2\kappa t \, .
\end{eqnarray}
Thus, we obtain the following result for the dressed solution:
\begin{eqnarray}
u_1(x,t) &=& \frac{2\Delta^*_p}{\Delta^2_p}\rme^{\rmi(\kappa^2 t - \delta)}\left[-\sinh(\kappa x - \theta)\cos\varphi
+ \rmi\sinh(\kappa x + \theta)\sin\varphi\right]\, ,\\
v_1(x,t) &=& 1 + \frac{2(2\rmi\kappa\alpha - 1)}{\Delta_p} + \frac{4\rmi\kappa\alpha(\rmi\kappa\alpha - 1)}{\Delta^2_p}\, ,\\
\Delta_p & = & \cosh^2(\kappa x - \theta)\cos^2\varphi + \cosh^2(\kappa x + \theta)\sin^2\varphi
- \rmi\kappa (x\sinh2\theta\sin2\varphi - 2\kappa t) \; .\nonumber
\end{eqnarray}
The result we obtained coincides with the one found in \cite{side9}. 

Let us consider now the case when $\epsilon = -1$. Relation (\ref{f0sq}) is rewritten as
\begin{equation}
|F_{0}^{3}|^2 - |F_{0}^{1}|^2 = 1
\end{equation}
hinting at the following parametrization:
\[F_{0}^{1} = \sinh\theta \rme^{\rmi(\delta + \varphi)},\qquad F_{0}^{3} = \cosh\theta\rme^{\rmi(\delta - \varphi)},
\quad \theta >0 \, .\]
Then we have
\begin{eqnarray}
|F^1(x,t)|^2 &=& \sinh^2(\kappa x - \theta)\cos^2\varphi + \sinh^2(\kappa x + \theta)\sin^2\varphi \, ,\\
\alpha(x,t) &=& x\sinh2\theta\sin2\varphi + 2\kappa t 
\end{eqnarray}
and substituting into (\ref{u1_doubl}) and (\ref{v1_doubl}) we obtain:
\begin{eqnarray}
u_1(x,t) &=& \frac{2\Delta^*_m}{\Delta^2_m}\rme^{\rmi(\kappa^2 t - \delta)}\left[\cosh(\kappa x - \theta)\cos\varphi
+ \rmi\cosh(\kappa x + \theta)\sin\varphi\right],\label{u1_doubl3}\\
v_1(x,t) &=& 1 + \frac{2(2\rmi\kappa\alpha + 1)}{\Delta_m} + \frac{4\rmi\kappa\alpha(\rmi\kappa\alpha + 1)}{\Delta^2_m}\, ,
\label{v1_doubl3}\\
\Delta_m & = & \sinh^2(\kappa x - \theta)\cos^2\varphi + \sinh^2(\kappa x + \theta)\sin^2\varphi
- \rmi\kappa (x\sinh2\theta\sin2\varphi + 2\kappa t) \;. \nonumber
\end{eqnarray}
It is not hard to see that $\Delta_m$ could vanish for particular values of the parameters, thus we obtain
singular solutions.
\end{enumerate}

Unlike the quadruplet solutions in previous subsection, the soliton-like solutions we derived here are related to a pair
of discrete eigenvalues (doublet) of the scattering operator. This is the reason why we call such solutions "doublet" solutions.

\subsection{Degenerate Case}\label{ratio}

Hereafter, we shall consider the case when the poles of the dressing factor are all real. Like in
doublet case (see previous subsection), the dressing factor
\begin{equation}
\mathcal{G}(x,t,\lambda) = \openone + \lambda\sum_{i=1}\left[\frac{B_i(x,t)}{\mu_i(\lambda - \mu_i)}
+ \frac{HB_i(x,t)H}{\mu_i(\lambda + \mu_i)}\right] ,\qquad \mu_i\in\bbr\backslash\{0\}
\end{equation}
and its inverse
\begin{equation}
\hat{\mathcal{G}}(x,t,\lambda) = \openone + \lambda\sum_{i=1}\left[\frac{Q_{\epsilon}B^{\dag}_i(x,t)Q_{\epsilon}}
{\mu_i(\lambda - \mu_i)} + \frac{Q_{\epsilon}HB^{\dag}_i(x,t)HQ_{\epsilon}}
{\mu_i(\lambda + \mu_i)}\right] \;
\end{equation}
have the same poles. Then the identity $\mathcal{G}(\lambda)\hat{\mathcal{G}}(\lambda) = \openone$ gives
rise to the following algebraic relations:
\begin{eqnarray}
&& B_iQ_{\epsilon}B^{\dag}_i = 0\; ,\label{algrel1_r}\\ 
&& \tilde{\Omega}_i Q_{\epsilon}B^{\dag}_i + B_iQ_{\epsilon}\tilde{\Omega}^{\dag}_i = 0
\label{algrel2_r}
\end{eqnarray}
where 
\begin{equation}
\tilde{\Omega}_i = \openone + \frac{B_i}{\mu_i} +
\sum_{j\neq i}\frac{\mu_iB_j}{\mu_j(\mu_i - \mu_j)} +
\sum_{j} \frac{\mu_iHB_jH}{\mu_j(\mu_i + \mu_j)}\; .
\label{Omega_i}
\end{equation}
Like in the previous case, relation (\ref{algrel1_r}) could be reduced to
\begin{equation}
F^T_iQ_{\epsilon}F^*_i = 0
\label{fquad}
\end{equation}
for the vector $F_i$ involved in the decomposition (\ref{b_fac}). It is clear that for
$Q_{\epsilon} = \openone$ (\ref{fquad}) leads to $F_i = 0$ and $\mathcal{G}$ becomes equal
to $\openone$. A nontrivial result is possible only for pseudo-Hermitian reduction, i.e.
$Q_{\epsilon} = \diag(1,-1,1)$. 

From this point on, we shall be interested in the case when $X_i$ and $F_i$ are just
$3$-vectors. Relation (\ref{algrel2_r}) implies there exists a scalar function $\beta_i$ such
that
\begin{equation}
\tilde{\Omega}_i Q_{\epsilon} F^*_i = X_i\beta_i\, , \qquad \beta^*_i	= - \beta_i\;.
\label{algrel2_ra}
\end{equation}
Relation (\ref{algrel2_ra}) is a linear system of equations for $X_i$. That system
could easily be solved when $\mathcal{G}$ has a single pair of real poles. In this case for $X$
we have:
\begin{equation}
X = \left(\beta - \frac{F^THQ_{\epsilon}F^*}{2\mu}\,H\right)^{-1}Q_{\epsilon} F^*\; .
\label{fx_sys3}\end{equation} 

In order to obtain $F_i$ and $\beta_i$, one considers the differential equations satisfied
by the dressing factor, see \eqref{g_pde1} and \eqref{g_pde2}. Analysis rather similar to that
in previous subsections shows that $F_i$ depend on bare fundamental solution according to
formula \eqref{f_psi0} while $\beta_i$ is determined by:
\begin{equation}
\beta_i = \beta_{i,0} - F^T_{i,0}\hat{\psi}_0(x,\mu_i)
\partial_{\lambda}{\psi}_0(x,\mu_i)\tilde{K}_{\psi_0}(\mu_i)Q_{\epsilon}F^*_{i,0}	
\end{equation}
where $\tilde{K}_{\psi}(\lambda) = \hat{\psi}(\lambda)Q_{\epsilon}\hat{\psi}^{\dag}(\lambda^*)Q_{\epsilon}$ 
measures the "deviation from invariance" of the solution $\psi$ and $\beta_{i,0}$ is an imaginary
$x$-independent scalar function. Like in the previous cases, \eqref{g_pde2} leads to the substitution rule 
(\ref{f_j0_evol}) recovering the time dependence of $F_{i}$. The time dependence of $\beta_i$
is obtained through the following substitution rule:
\begin{equation}
\beta_{i,0}  \to  \beta_{i,0} - \rmi F^T_{i,0}\left. 
\frac{\rmd f(\lambda)}{\rmd\lambda}\right|_{\lambda = \mu_i}\tilde{K}_{\psi_0}(\mu_i) Q_{\epsilon}F^*_{i,0}\, t
\label{t_recov2}
\end{equation}
where $f(\lambda)$ is the dispersion law. 

We shall focus now on the simplest case when the dressing factor has one pair of real poles. Assume
the seed solution $S^{(0)}$ and the corresponding fundamental solution $\psi_0$ are given by
(\ref{seed}) and \eqref{psi_0} respectively. Due to \eqref{fquad}, for the components of the 3-vector
$F_0 = (F_{0}^{1}, F_{0}^{2}, F_{0}^{3})$ we have
\begin{equation}
|F_{0}^{1}|^2	+ |F_{0}^{3}|^2 = |F_{0}^{2}|^2\; .
\label{fquad2}\end{equation}
Without any loss of generality we could set $F_{0}^{2} = 1$. Then the form of (\ref{fquad2})
suggests the parametrization:
\[ F_{0}^{1}(t) = \cos\theta\,\rme^{\rmi\left(\frac{\mu^2t}{3}+\delta + \varphi\right)}, \qquad F_{0}^{3}(t) = \sin\theta\, \rme^{\rmi\left(\frac{\mu^2t}{3}+\delta - \varphi\right)}, \qquad \theta\in [0, \pi/2]\; .\]
Then $F$ could be written down as:
\begin{equation}
F(x,t) = \left(\begin{array}{c}
\rme^{\rmi\left(\frac{\mu^2t}{3} + \delta\right)}\left[\rme^{\rmi\varphi}\cos\mu x\cos\theta
+ \rmi\,\rme^{-\rmi\varphi}\sin\mu x\sin\theta\right] \\
\rme^{-\frac{2\rmi\mu^2t}{3}}\\
\rme^{\rmi\left(\frac{\mu^2t}{3} + \delta\right)}\left[\rme^{-\rmi\varphi}\cos\mu x\sin\theta
+ \rmi\,\rme^{\rmi\varphi}\sin\mu x\cos\theta\right]
\end{array}\right)	.
\label{f_deg}\end{equation}
A relatively simple computation shows that $\beta$ depends linearly on $x$ and $t$
as follows:
\begin{equation}
\beta = \rmi (2\mu t + x\sin 2\theta\cos2\varphi)\, .
\label{alpha}\end{equation}
Thus, taking into account formula \eqref{s10_eq}, dressed solution acquires the form:
\begin{eqnarray}
u_1(x,t) &=&  - \frac{2\left(\mu\beta + |F^1|^2\right)F^2\left(F^3\right)^*}{\left(\mu\beta - |F^1|^2\right)^2}\, ,
\label{u1_5sym}\\
v_1(x,t) &=& 	\frac{\left(\mu\beta + |F^1|^2\right)\left(\mu\beta - 1 - |F^3|^2\right)}
{\left(\mu\beta - |F^1|^2\right)^2}\, .\label{v1_5sym}
\end{eqnarray}
It could easily be checked that (\ref{u1_5sym}) and (\ref{v1_5sym}) satisfy the constraint
(\ref{constr}) provided $\alpha$ and $F$ obey (\ref{fquad}) but are otherwise arbitrary.

After substituting (\ref{f_deg}) and (\ref{alpha}) into (\ref{u1_5sym}) and (\ref{v1_5sym})
and making some further simplifications, we derive 
\begin{eqnarray}
&&u_1(x,t)=\nonumber \\ 
&& -\frac{2\Delta_d^*}{\Delta_d^2}\rme^{-\left(\rmi\mu^2t + \delta + \frac{\pi}{4}\right)}
\left[\cos\left(\varphi + \frac{\pi}{4}\right)\sin(\mu x + \theta)
- \rmi\,\sin\left(\varphi + \frac{\pi}{4}\right)\sin(\mu x - \theta)\right]\label{u1},\label{u1_ratio}\\
&&v_1(x,t)=\frac{\Delta_d^*(\Delta_d^* - 2)}{\Delta_d^2}\label{v1_ratio}
\end{eqnarray}
where
\[\Delta_d = \cos^2(\mu x + \theta)\cos^2\left(\varphi + \frac{\pi}{4}\right) + \cos^2(\mu x - \theta)
\sin^2 \left(\varphi + \frac{\pi}{4}\right) - \rmi\mu(2\mu t + x\sin2\theta\cos 2\varphi).\]
The solutions (\ref{u1_ratio}) and (\ref{v1_ratio}) contain terms that are linear in $x$ and $t$ as well as terms
that are bounded oscillating functions. This is why we call such solutions quasi-rational. It is easy to see that
the denominator $\Delta_d$ has zeros for particular values of the parameters. For example, when $\theta = \pi/2$
it is zero at $x=0$ and $t=0$.

\section{Conclusion}\label{con}

We have shown in that paper how Zakharov-Shabat's dressing method can be applied to the linear bundle
(\ref{ghf_lax1}) whose potential functions are subject to the boundary condition (\ref{triv_back}).
Following the general algorithm described in Section \ref{dressing}, we have constructed special
solutions to the generalized Heisenberg ferromagnet equation (\ref{ghf}). The simplest class of
such solutions corresponds to a dressing factor with simple poles, see (\ref{dress_fac}). We have
seen that the location of the poles in $\lambda$-plane affects the form of solutions. There exist
three "pure" cases: the poles are complex numbers in generic position, the poles are imaginary and the
poles are real. The first two options lead to soliton-like solutions of quadruplet and doublet type
respectively, see formulas (\ref{u1_1}), (\ref{v1_1}), (\ref{u1_2}), (\ref{v1_2}), (\ref{u1_3}), (\ref{v1_3}),
(\ref{u1_doubl2}), (\ref{v1_doubl2}), (\ref{u1_doubl3}) and (\ref{v1_doubl3}). All the quadruplet solutions
we have derived are non-singular while the doublet solutions could have singularities in the pseudo-Hermitian case.

The case of real poles differs much from the others. It introduces certain degeneracy in the spectrum
of the scattering operator and leads to quasi-rational solutions, see (\ref{u1_ratio}) and (\ref{v1_ratio}).
We have seen such degeneracy is possible only if we have a pseudo-Hermitian reduction. This is a rather
essential difference between the Hermitian and pseudo-Hermitian reductions. The quasi-rational solutions
(\ref{u1_ratio}) and (\ref{v1_ratio}) could be non-singular for particular values of the parameters.

Apart from the "pure" cases we have discussed in the main text, there exists a situation when some poles
of the dressing factor are generic complex numbers while the rest are either real or imaginary numbers.
Clearly, the analysis of this "mixed" case can be reduced to the considerations we have already demonstrated. 

Though we have discussed in detail special solutions to (\ref{ghf}), similar procedures could be used to
derive explicit solutions to any NLEE belonging to the integrable hierarchy of (\ref{ghf_lax1}). The
solutions of such a NLEE will have the same $x$-dependence as those of (\ref{ghf}) but a different
$t$-dependence. 

In the present paper, we have focused on the simplest class of solutions obeying the asymptotic behavior
(\ref{triv_back}). A possible way of extending our results is by looking for solutions having more complicated
behavior, e.g. non-trivial background solutions or periodic solutions. In the last few decades, nontrivial
background solutions have become a topic of increased interest due to the connection to phenomena like rogue
(freak) waves, see \cite{per,ahm1,deg1,rud2}. Such solutions were obtained for classical integrable equations
like (scalar) nonlinear Schr\"odinger equation \cite{per, ahm1,ahm2,hone}, $3$-wave equation \cite{deg1,deg2}
and for scalar derivative nonlinear Schr\"odinger equation \cite{dnlsratio, multisoldnls}. This is why it is
interesting to explicitly construct similar type solutions for the generalized Heisenberg equation and find out
how the properties of (\ref{ghf}) and its spectral problem will change if $u$ and $v$ have different behavior.

Another direction to extend our results is by studying auxiliary spectral problems and the corresponding integrable
hierarchies of NLEEs for other symmetric spaces and other reductions. This includes the study of rational bundles like 
\[ L(\lambda) = \rmi\partial_x - \lambda S_1 - \frac{1}{\lambda} S_{-1},\qquad \lambda\in\bbc\backslash\{0\}\, ,\]
where
\[S_{\pm 1} = \left(\begin{array}{ccc}
0 & \pm u & v\\
\pm \epsilon_1 u^* & 0 & 0\\
\epsilon_2 v^* & 0 & 0\end{array}\right),\qquad
\epsilon_1^2 = \epsilon_2^2=1 \, .\]
The above rational bundle can be viewed as a deformation of (\ref{ghf_lax1}) and it has much more complicated
properties than the linear bundle we have considered here, see \cite{gmv,tmf} for a discussion of the Hermitian
case ($\epsilon_1 = \epsilon_2 = 1$). We intend to discuss all these issues in more detail elsewhere.

\section*{Acknowledgments}
The work has been supported by the NRF incentive grant of South Africa and grant DN 02-5 of
Bulgarian Fund "Scientific Research".

\end{document}